\documentclass[conference]{IEEEtran}
\usepackage[pass]{geometry}

\usepackage{cite}
\usepackage{amsmath,amssymb,amsfonts}

\usepackage[longend,ruled,vlined]{algorithm2e} 
\SetKwIF{If}{ElseIf}{Else}{if}{}{else if}{else}{end if}
\let\oldnl\nl
\newcommand{\nonl}{\renewcommand{\nl}{\let\nl\oldnl}}

\usepackage{setspace}
\usepackage{subfigure}
\usepackage{caption}
\usepackage{graphicx}
\usepackage{textcomp}
\usepackage{xcolor}
\usepackage[bookmarks=false]{hyperref}
\usepackage{url}  
\usepackage{tabularx} 
\usepackage{booktabs}

\usepackage{multirow}
\usepackage{array}
\newcolumntype{P}[1]{>{\centering\arraybackslash}p{#1}}
\newcolumntype{M}[1]{>{\centering\arraybackslash}m{#1}}
\usepackage{longtable}
\usepackage{xtab,afterpage}
\usepackage{comment}
\usepackage{float}
\usepackage{listings}
\usepackage{upgreek}
\usepackage{enumitem}
\usepackage[official]{eurosym}
\usepackage{ragged2e}

\def\BibTeX{{\rm B\kern-.05em{\sc i\kern-.025em b}\kern-.08em
    T\kern-.1667em\lower.7ex\hbox{E}\kern-.125emX}}

\title{An Isolation-aware Online Virtual Network Embedding via Deep Reinforcement Learning}

\author{
\IEEEauthorblockN{Ali Gohar\IEEEauthorrefmark{1}\IEEEauthorrefmark{2}, Chunming Rong\IEEEauthorrefmark{2}, Sanghwan Lee\IEEEauthorrefmark{3}}
\IEEEauthorblockA{\IEEEauthorrefmark{2}Department of Electrical Engineering and Computer Science, University of Stavanger, Stavanger, Norway\\
\IEEEauthorrefmark{3} College of Computer Science, Kookmin University, Seoul, Korea \\
 \IEEEauthorrefmark{1}Corresponding author: Ali Gohar\\
    \{ali.gohar, chunming.rong\}@uis.no, sanghwan@kookmin.ac.kr}

}

\begin{document} 

\maketitle

\let\thefootnote\relax\footnotetext{This paper has been accepted in 2023 IEEE/ACM 23rd International Symposium on Cluster, Cloud and Internet Computing Workshops (CCGridW).}

\begin{abstract}
	
	Virtualization technologies are the foundation of modern ICT infrastructure, enabling service providers to create dedicated virtual networks (VNs) that can support a wide range of smart city applications. These VNs continuously generate massive amounts of data, necessitating stringent reliability and security requirements. In virtualized network environments, however, multiple VNs may coexist on the same physical infrastructure and, if not properly isolated, may interfere with or provide unauthorized access to one another. The former causes performance degradation, while the latter compromises the security of VNs. Service assurance for infrastructure providers becomes significantly more complicated when a specific VN violates the isolation requirement.

	In an effort to address the isolation issue, this paper proposes isolation during virtual network embedding (VNE), the procedure of allocating VNs onto physical infrastructure. We define a simple abstracted concept of isolation levels to capture the variations in isolation requirements and then formulate isolation-aware VNE as an optimization problem with resource and isolation constraints. A deep reinforcement learning (DRL)-based VNE algorithm ISO-DRL\_VNE, is proposed that considers resource and isolation constraints and is compared to the existing three state-of-the-art algorithms: NodeRank, Global Resource Capacity (GRC), and Mote-Carlo Tree Search (MCTS). Evaluation results show that the ISO-DRL\_VNE algorithm outperforms others in acceptance ratio, long-term average revenue, and long-term average revenue-to-cost ratio by 6\%, 13\%, and 15\%.

\end{abstract}


\section{Introduction}
\label{intro}

Network Function Virtualization (NFV) technology enables network operators to deploy, scale, and manage Virtual Network Functions (VNFs) more flexibly and efficiently. NFV technology is frequently used in conjunction with network slicing to support the diverse needs of 5G mobile networks. Network slicing allows partitioning a physical network infrastructure into multiple logical, end-to-end, isolated Virtual Networks (VNs), each with its own characteristics and performance parameters. A VN can be composed of a series of VNFs, and if a set of VNFs is chained in the order in which traffic flows through the network, it is called a Service Function Chain (SFC). This novel approach in the mobile network ecosystem allows for a role separation between Infrastructure Providers (InP) and Service Providers (SPs) \cite{8057045}. 

To create VNs, InP deploys the necessary VNFs on the InP infrastructure, which comprises heterogeneous resources such as computing, storage, and network, also known as a Substrate Network (SN). Efficient mapping of the VN towards the SN primarily deals with the VNF placement or is a typical Virtual Network Embedding (VNE) problem. VNE has been extensively researched in the literature from various perspectives, including resource allocation, power consumption, and security \cite{SUN2022103361}. Due to its NP-Completeness \cite{Andersen2002TheoreticalAT}, various solutions for solving the VNE problem have been proposed and are broadly classified into the following categories \cite{8968720} \cite{SUN2022103361}; 

(i) Mathematical optimization-based: \cite{5951812} model the VNE problem using MIP, but to obtain the solution in polynomial time, a rounding-based approach is used to relax the integer constraints. \cite{8057036} use ILP to model the VNE problem as a set cover problem and propose algorithms with a logarithmic approximation factor to place the VNFs. Due to the high computational complexity, these offline approaches are challenging to apply to online placement scenarios; 

ii) Heuristic-based:  To increase the long-term average revenue and acceptance ratio, a PageRank-inspired VN embedding algorithm is proposed to compute ranking values using the Random Walk (RW) in \cite{10.1145/1971162.1971168}. \cite{6847918} propose an algorithm based on the Global Resource Capacity (GRC) metric to maximize the revenue-to-cost ratio. \cite{LIU2015151} presents an abstracted numerical concept of security levels and demands followed by the formulating a security-aware VNE problem that considers takes security constraints. These algorithms cannot adapt to a various environments and may result in local optimal solutions.

iii) Reinforcement learning (RL)-based: \cite{7859375} formalize the VNE problem using the Markov Decision Process (MDP) and use the Monte-Carlo Tree Search (MCTS) algorithm to maximize the revenue-to-cost and acceptance ratio. To minimize power consumption, \cite{8945291} employs a policy gradient approach combined with the Lagrange relaxation technique. Their evaluations, however, are limited to using a grid or star topology to represent the SN. The RL algorithms offer significant advantages over traditional methods, including adaptation to changing network conditions, optimization for a global solution, handling uncertainty, and minimizing human intervention through continual learning.

SPs use a variety of third-party InP SNs for data computation and/or storage, which leads to inherent confidentiality, integrity, and availability vulnerabilities for VNs \cite{8945291}. However, as an essential dimension, researchers have only discussed VN isolation in this context from theoretical perspectives and challenges that VNs face concerning isolation, which is the property that one VN operates independently of other VNs using the same infrastructure \cite{9200939}. 

In this paper, we present an abstracted numerical concept of isolation levels and requirements, followed by the formulation of an isolation-aware VNE problem that considers isolation constraints. Furthermore, we propose a Deep RL (DRL)-based algorithm ISO-DRL\_VNE for adaptive VN placement. ISO-DRL\_VNE maximizes long-term average revenue by learning to make VN placement decisions based on past decision performance rather than a hypothetical environment. The main contributions of this paper are:

\begin{itemize}
	\item a concept of abstractions to represent isolation levels and requirements,
	\item a model and description of the VNE problem constrained by isolation,
	\item a solution based on DRL for adaptive VN placement,	
	\item a performance comparison of the proposed solution with three existing works: NodeRank \cite{10.1145/1971162.1971168}, GRC \cite{6847918}, and MCTS \cite{7859375}.         	      	          
\end{itemize}

The rest of this paper is organized as follows: Section \ref{sec:isolation} presents isolation levels. Section \ref{description} describes the problem model. Section \ref{algorithm} describes the solution algorithm. Section \ref{evaluation} compares the performance of the proposed algorithm with the three existing algorithms, and finally, Section \ref{conclusion} concludes this study.

\section{Isolation-aware Virtual Network Embedding} \label{sec:isolation}

The 3GPP highlights several important issues that need to be addressed to maintain an isolated environment for future mobile networks \cite{3gpp_tr_23.799}. Previous studies have discussed the challenges of achieving end-to-end isolation, identifying enabling technologies and the need for standardized approaches for designing isolated VNs or slices \cite{BARAKABITZE2020106984} \cite{kotulski_2018}. The isolation types and parameters are further explored in \cite{kotulski_2018}. It is recognized that defining the proper isolation parameters and points is a critical issue \cite{kotulski_2018}. Based on available technologies and current literature, this paper identifies three major isolation points that can be implemented to ensure VN and SN separation, as follows (i) isolation between different SN nodes, (ii) VN nodes deployed on a SN node, and (iii)  VN nodes co-deployed on the same SN node.

To address VN node isolation issues, each SN node is assumed to have an isolation level that allows each VN node on the shared infrastructure to operate independently of other VN nodes. The variation in isolation levels stems from the realization that, depending on the service type, not all VN nodes require the same isolation level. We identify the following abstracted isolation between an SN node and a VN node based on the assumptions stated above.

\begin{itemize}
	\item  The isolation level of a SN node should be equal to or greater than the isolation requirement level of the VN node deployed on the SN node. 
	\item  The isolation level of a VN node should be equal to or less than the isolation level of the SN node hosting the VNF. 
	\item  The isolation level of a VNF should be equal to or less than the isolation requirement level of the VNFs co-deployed on the same SN node.
\end{itemize}

\section{Problem model and description}
\label{description}

Two actors are considered in the VNE problem: InPs and SP. The InP receives a VN Request (VNR) from SP representing a specific VN and must be allocated on SN. The following two subsections describe the types of resources provided by InP SN and the requirements for VNRs.

\subsection{SN Model}
The SN is composed of the following two elements:
\begin{enumerate}
	\item \emph{Computation platforms (CPs):}
	      We indicate $N^{S}$ as the set of CPs where $A_{N}^{S}$ represents a configuration for each CP. For each CP configuration, we consider the following resources:
	      \begin{itemize}
	      	\item the amount of processing $CPU(n^{s})$ in [vCPU];
	      	\item isolation level available $ISA(n^{s})$ in [Units]; $\forall$ $n^{s} \in N^{S}$.
	      \end{itemize}
	      	      	      	        
	\item \emph{Logical connections (LCs):} The LCs connect CPs and are provided by network operators. We indicate
	      $L^{S}$ as the set of LCs where $A_{L}^{S}$ represents a configuration for each LC. We consider the following resources for each LC configuration:
	      \begin{itemize}
	      	\item the amount of data rate $BW(l^{s})$ in [Mb/s]; $\forall$ $l^{s} \in L^{S}$
	      \end{itemize}
\end{enumerate}

The SN model can now be formalized using graph theory and can be represented as a weighted undirected graph by $G^{S}= \{N^{S},L^{S}, A_{N}^{S}, A_{L}^{S}\}$.
	      		      	
\subsection{VNR Model}
	      		      		      	
A VNR represents a specific VN provided by SP for its users, and has the following requirements:

\begin{enumerate}
	\item \emph{Computing requirements:}  We indicate $N^{V}$ as the set of virtual nodes (i.e., VNFs) where $A_{V}^{S}$ represents a configuration for each virtual node. For each virtual node configuration, we consider the following resources:
	      \begin{itemize}[leftmargin=*]	       
	      	\item computing resource requirement $CPU(n^{v} )$, in [vCPU];
	      	      	      	     
	      	\item isolation level requirement $ISR(n^{v})$ in [Units]; $\forall$ $n^{v} \in N^{V}$.
	      \end{itemize}
	      	      	      	      	      	      	      	         	 
	\item \emph{Networking requirement}: Since, virtual nodes are connected together. We consider virtual links that connect various virtual nodes. We indicate
	      $L^{V}$ as the set of virtual links where $A_{L}^{V}$ represents a configuration for each virtual link. We consider the following resources for each virtual link configuration:
	      \begin{itemize}
	      	\item data rate requirement  $BW(l^{v})$ in [Mb/s]; $\forall$ $l^{v} \in L^{V}$   
	      	      	      	    
	      \end{itemize}	      	      	      	      	      	      	     		
\end{enumerate}

The VNR model is formalized using graph theory and is represented as $G^{V}=\{N^{V},L^{V}, A_{N}^{V}, A_{L}^{V}\}$. 

\subsection{VNE Problem Description}

\begin{figure}[!htb]
	\centering
	\includegraphics[width=0.45\textwidth]{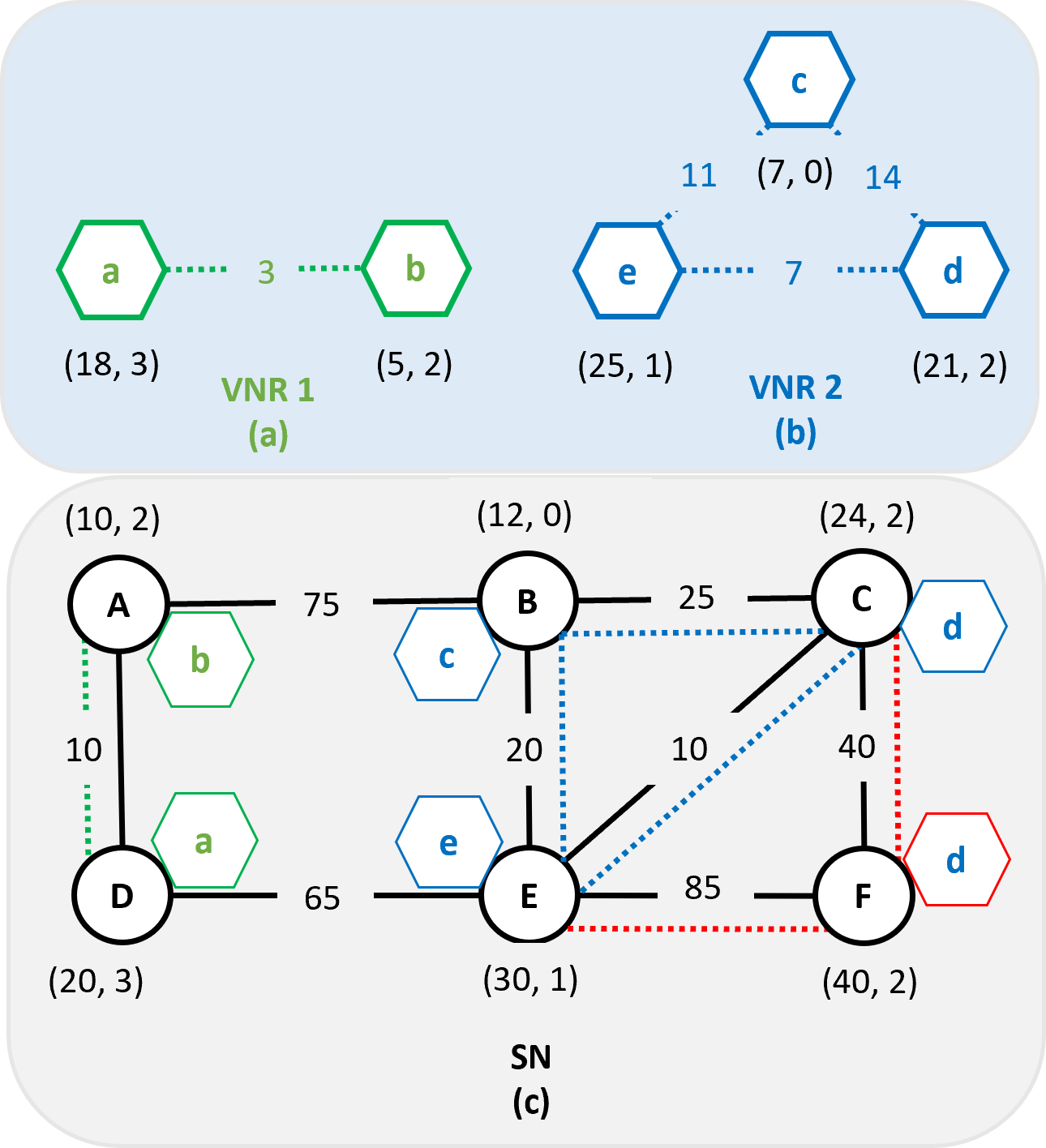}
	\caption{SN \& VN model with possible embedding examples}
	\label{fig:vne_problem_description}
\end{figure}

The VNE problem can be express as mapping $ G^{V}$ to $G^{V}(N^{V}, L^{V}) \xrightarrow[]{} G^{S}(N^{S_{i}}, L^{S_{i}})$ where $N^{S_{i}} \in N^{S}, L^{S_{i}} \in L^{S}$. Figure \ref{fig:vne_problem_description} shows two different embedding scenarios of VNRs. Moreover, (a) and (b) depict two distinct VNRs, (VNR 1 and VNR 2), while (c) depicts an SN. A Hexagon represents the VN nodes, and a dotted line represents their VL demand. 
A circle represents an SN node, and a solid line connecting two circles represents an SN link. The numbers in brackets next to the virtual nodes indicate the computing and isolation level requirements for two VNRs. The rest numbers on those links indicate the bandwidth needs of the virtual links. Similarly, the numbers in brackets in SN represent the available computing resources, and isolation levels, respectively. The rest numbers represent the available bandwidth resources.
  
We highlight three possible scenarios in which VNRs are embedded in the SN. The virtual nodes a and b in VNR 1 are mapped to the A and D. However, the virtual nodes c, d, and e in VNR 2 can be mapped to the B, C, and E or the C, E, and F, respectively. It should be noted that the virtual node d of VNR 2 can be embedded in either C or F; however, embedding virtual node d of VNR 2 in F will consume more SN node and link resources and is not an optimal allocation.

\subsection{Evaluation Metrics}

This paper assumes that the InP charges SPs by the “pay-as-you-go” pricing model. As a result, the revenue generated by an InP accepting a VNR is defined as follows:
  
\begin{equation}
	REV (G^{V}) = \alpha
	\sum ^{|N^{V}|}_{i=1} CPU (n_{i}^{v}) + \beta \sum ^{\left| L^{V}\right| }_{j=1} BW (l_{j}^{v})
	\label{eq:revenue}
\end{equation}
	      	      	      	      
Eq.~\ref{eq:revenue}, shows the  revenue of InP, where $CPU (n_{i}^{v}) $ represents the computing resources occupied by the virtual node $(n_{i}^{v})$. $BW (l_{j}^{v})$ represents the bandwidth resource occupied by the virtual link $l_{i}^{v}$. where $\alpha$ is the unit price of resource for SN nodes (CPs),
and $\beta$ is the unit price of bandwidth for SN links (LCs)

Eq.~\ref{eq:cost}, shows the cost to InP for providing SN resource services to the SP and is formulated as follows:
\begin{equation}
	CST(G^{V})=\sum ^{|N^{V}|}_{i=1} CPU (n_{i}^{v}) + \sum ^{| L^{V}| }_{j=1}\sum ^{| L^{S}| }_{k=1} BW (l_{jk}^{vs})
	\label{eq:cost}
\end{equation}
where $CPU (n_{i}^{v}) $ represents the computing resources occupied by the virtual node $(n_{i}^{v})$ after successful embedding. $BW (l_{jk}^{vs})$ represents the bandwidth resource occupied by the virtual link $l_{jk}^{vs}$ with link splitting.

VNRs arrive with varying time distributions and resource requirements. It is preferable to have a VNE algorithm that accepts more VNRs. The ratio of accepted VNRs to the total number of VNR that arrived is defined as:
     	      	
\begin{equation}
	ACR=\lim _{T\rightarrow \infty }\dfrac{\sum ^{T}_{t=0}accepted\left( G^{V}, t \right) }{\sum ^{T}_{t=0}arrive \left( G^{V}, t \right) }
	\label{eq:acceptance_ratio}
\end{equation}

Eq.~\ref{eq:acceptance_ratio} shows the acceptance rate where $accepted(G^{V}, t)$ denotes the total number of VNRs that are successfully embedded and $arrive(G^{V}, t)$ represents the total number of VNRs that make resource requests. 

The long-term average revenue (LAR), is used to assess the overall effectiveness of a VNE algorithm, is defined as:

\begin{equation}
	LAR=\lim _{T\rightarrow \infty }\dfrac{1}{T}\sum ^{T}_{t=0} REV\left(G^{V}, t \right)
	\label{eq:long_term_average_revenue}
\end{equation}

Eq.~\ref{eq:long_term_average_revenue} shows $(G^{V}, t)$ the revenue of successful VNR allocation arrived at time $t$ where $T$ is the total elapsed time. A higher LAR leads to a higher profit for the InP.

A higher long-term average revenue-to-cost ratio (LRC) indicates a higher resource utilization efficiency to meet the resource needs of more VNRs. LRC is defined in Eq.~\ref{eq:revenue_2_cost_ratio} as:   
\begin{equation}
	LRC=\lim _{T\rightarrow \infty }\dfrac{\sum ^{T}_{t=0}REV\left( G^{V}, t\right) }{\sum ^{T}_{t=0}CST\left( G^{V}, t\right) }
	\label{eq:revenue_2_cost_ratio}
\end{equation}

The VNE process needs to meet the following constraints:

\begin{equation}
	R_{CPU}(n^{s})= CPU\left( n^{s}\right) - \sum ^{|VNR|}_{i=1, n^{v}\rightarrow n^{s} } CPU\left( n_{i}^{v}\right)
	\label{eq:remaining_cpu}
\end{equation}

\begin{equation}
	R_{BW}(l^{s})= BW(l^{s}) - \sum^{|VNR|}_{i=1, l^{v}\rightarrow l^{s} } BW( l_{i}^{v})
	\label{eq:remaining_bw}
\end{equation}

\begin{equation}
	R_{cpu}(n^{s}) \geq CPU(n^{v})
	\label{eq:constrain_cpu}
\end{equation}

\begin{equation}
	R_{BW}(l^{s}) \geq BW(l^{v})
	\label{eq:constrain_bw}
\end{equation}

\begin{equation}
	ISA(n^{s}) \geq ISR(n^{v})
	\label{eq:constrain_isolation}
\end{equation}

Eq.~\ref{eq:remaining_cpu} and Eq.~\ref{eq:remaining_bw} represent the remaining computing and bandwidth resources for the SN nodes and SN links. Eq.~\ref{eq:constrain_cpu} and Eq.~\ref{eq:constrain_bw} represent the computing and bandwidth constraints of a VNR.
Eq.~\ref{eq:constrain_isolation} represent the isolation constraints of VNR. Eq.~\ref{eq:constrain_cpu} and Eq.~\ref{eq:constrain_isolation} are followed during the node embedding stage, and Eq.~\ref{eq:constrain_bw} is followed during the link embedding stage.

In this paper, ACR, LAR, and LRC are all optimized together. The goal is to maximize following joint optimization, which is defined as:

\begin{equation}
	maximize \ Objective = \ E_{1} + E_{2} + E_{3}
\end{equation}
$where$ $E_{1} = ACR,\ E_{2} = LAR,\ E_{3} = LRC. $
\section{EMBEDDING ALGORITHM}
\label{algorithm}

The basic elements of RL are agent, environment, state, action, and reward. The Markov Decision Process (MDP) is used to describe the environment in RL. We model the placement of each VNR as a finite-horizon MDP in which the agent selects SN nodes to place VNFs of VNR at discrete time step $t \in (1,2,3,\ldots, T)$. We define three key elements: state, action, and reward as follows:

\subsubsection{State}

To represent the state information of SN, the SN features are extracted. Extracting all SN features would be unrealistic due to the increased computational complexity, and insufficient feature extraction would result in an inadequate representation of the environment. For each SN node, the following features are extracted:

\begin{itemize}
    \item $CPU(n^{s})$: The remaining CPU capacity of an SN node has a significant impact on allocation.
    \item $DEG(n^{s})$: Degree represents the number of LC directly connected to the SN node. Higher degree results in easier path finding between other SN nodes.
    \item  $SUM(n^{s})$: Represents the sum of the bandwidth of all LC connected to each SN node. A higher SUM is more likely to complete the link embedding in this SN node.	                  
\end{itemize}

Following feature extraction, the values are normalized and represented as a feature vector. The above properties of the $i-th$ SN node are represented as a 3-dimensional vector, as follows:
\begin{equation}
   v_{i}=(CPU(n_i^s), DEG(n_i^s), SUM(n_i^s))^T
	\label{eq:}
\end{equation}

All SN node features are represented as vectors and are combined into a feature matrix $M$, which is provided as input to the agent and updated as the SN state changes.

\begin{equation}
     M = (v_1,v_2\ldots v_n)^T
\end{equation}
\\
The feature matrix $M$ is expressed as follows:

\begin{equation}
    \displaystyle \begin{aligned} \begin{bmatrix} \begin{smallmatrix} CPU(n_1^s) & DEG(n_1^s) & SUM(n_1^s) \\ CPU(n_2^s) & DEG(n_2^s) & SUM(n_2^s)  \\ \ldots & \ldots & \ldots  \\ 
    CPU(n_k^s) & DEG(n_k^s) & SUM(n_k^s) \\ \end{smallmatrix} \end{bmatrix} \end{aligned}
\end{equation}

The purpose of normalization is to accelerate the training process and enable the agent to converge quickly.

\subsubsection{Action}
 
The action at time $t$ selects an SN node from the set of all SN nodes whose available resources exceed the requirements requested by VNF.

\subsubsection{Reward}

A reward signal is required to help the agent learn better actions. As a reward, the $LRC$ generated by a successful VNR embedding is used; otherwise, the agent receives 0 as a reward. Reward based on LRC reflects the utilization of network resources. High LRC values indicate that the selected actions by the agent are effective in realizing high VNE revenue, while low LRC values necessitate adjustments to the actions by the agent. 

\subsection{Policy Network}

\begin{figure}[t]
	\centering
	\includegraphics[width=0.5\textwidth]{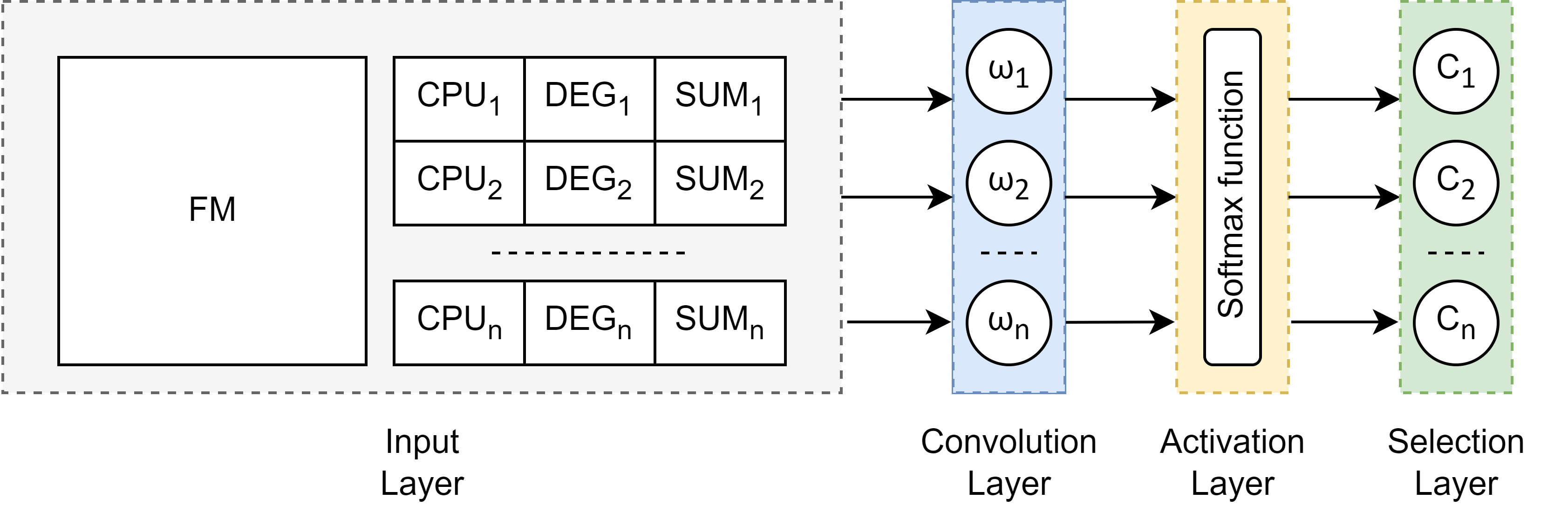}
	\caption{Policy network}
	\label{fig:policy_network}
\vspace{-0.5cm}
\end{figure}

In this paper, we use a four-layer ANN to construct a policy network as the learning agent, which takes the $M$ as input and outputs the probabilities of mapping VNR nodes to SN nodes. The policy network comprises four layers: the input layer, the convolution layer, the probability layer, and the selection layer as shown in Fig. \ref{fig:policy_network}.

The convolution layer receives a $M$ and transfers it to the input layer (CNL). CNL applies the convolution operation on $M$ to create a vector representing each SN node available resources and is given by Eq.~\ref{eq:convolution_operation}

\begin{equation}
	z_{i} = \omega.v_{{i}} + d
	\label{eq:convolution_operation}
\end{equation}

$z_{i}$ is the $i$th output of the convolution layer, $\omega$ is the weight of the convolution kernel vector, and $d$ is the deviation.  

Softmax is mainly used to measure the probability of each SN node based on the available resource in the probability layer and is given by Eq.~\ref{eq:softmax_operation}.

\begin{equation}
	P_{i}=\dfrac{e^{r_{i}}}{\sum k^{e^{r_{n}}}}
	\label{eq:softmax_operation}
\end{equation}

where $P_{i}$ represents the probability that the $i$th physical node is embedded.  $n$ is the number of feature vectors.
The selection layer filters the set of SN nodes with sufficient resource capacities, and the SN node with the highest probability is chosen for embedding.

\subsection{Training and Testing}

The gradient update method is used to learn policy network parameters. The gradient update procedure updates the loss function in the direction of gradient descent to optimize the parameters of the policy network. It optimizes both the reward mechanism and the policy network at the same time, and the result is as follows:

\begin{equation}
 \vspace{-0.1cm}
\mathcal{J}=\mathcal{J}-{\mathcal{J}}^{\prime}\times \alpha \times Reward
	\label{eq:reward}
\end{equation}

$\alpha$ is an important factor influencing the gradient update and training speed. A small value will slow down the training process. A large value will lead to poor decisions, and a value too large will produce unstable training. A batch updating approach is used to ensure learning stability and speed up updating the policy network parameters.

Algorithm \ref{algo:trainign_policy_network} shows the DRL-based training process for the VNE algorithm. All policy network parameters are initialized randomly in line 1 and trained for several epochs in line 2 during the training phase. Each VNR is mapped on SN in two stages.
In the first stage, lines 5-8 map the VNR nodes onto SN. The virtual links of VNR nodes are mapped using Breadth-First Search (BFS) in the second stage in line 10 if all VNR nodes are mapped.

The feature matrix ($M$) of SN is passed to the policy network in line 3 along with the VNR that is chosen from the $trainingSet$. Lists of potential SN nodes and the likelihood that they will map to VNR nodes are produced by the policy network in line 7. If no SN nodes or SN links with enough resources are present, the mapping fails. Line 8 includes the gradient calculation. Following the successful mapping of VNR nodes and links, line 11 determines the reward. Before calculating the gradient update in line 12, we provide $LRC$ as a $Reward$ which is used in Eq.~\ref{eq:reward}. 

During testing of the policy network, a greedy strategy is employed, selecting the SN node with the highest probability as the mapping node for the VNR virtual node. The testing process of the VNE algorithm based on DRL is demonstrated in Algorithm \ref{algo:testing_policy_network}.

\RestyleAlgo{ruled}
\DontPrintSemicolon
\LinesNumbered
\SetKwInOut{Input}{Input}
\SetKwInOut{Output}{Output}
\begin{algorithm}
	\renewcommand{\arraystretch}{1}
	\caption{Training Policy Network }
	\label{algo:trainign_policy_network}
	\Input{SN, trainingSet, epochNumber, $\alpha$, batchSize}
	\Output{Policy network parameters}
	{Initialize all parameters of the policy network }\;
	\While {$iterationNumber \leq epochNumber$ }{
		\For {VNR $\in$ trainingSet}{
			{$counter=0$}\;            
			\For {VNR\_node $\in$ VNR}{
				{receive feature matrix ($M$) }\;
				{get probability distribution of SN nodes }\;
				{compute the gradient }\;
			}
			\If { isMapped ($\forall$ VNR\_nodes $\in$ VNR)  }{
				\eIf { linkMappingUsingBFS(VNR)}{
					{reward = rewardCalculation(VNR) }\;
					{gradientUpdate(reward, $\alpha$) }\;
					}{
					{clearGradient()}\;
				}
				{$counter+=$1}\;  
				\If { $counter == $ batchSize}{
					{$counter == $ 0}\;		
				}
			}
		}
		{$iterationNumber+=1 $}\;
	}
 
\end{algorithm}

\begin{algorithm}

	\renewcommand{\arraystretch}{1} 
	\caption{Testing Policy Network }
	\label{algo:testing_policy_network}
	\Input{SN, testingSet}
	\Output{Results for three evaluation objectives}
	{Initialize all policy network parameters}\;
		\For {VNR $\in$ testingSet}{
			\For {VNR\_node $\in$ VNR}{
				{get feature matrix ($M$) }\;
				{get mapping probability of SN node }\;
			}
			\If { isMapped ( $\forall$ VNR\_nodes $\in$ VNR )  }{
				\If { linkMappingUsingBFS(VNR)}{
                    return $embeddingSuccess$ 
				}
			}
		} 
\end{algorithm}

\begin{table}[!htb]
	\caption{Evaluation parameters}
	\vspace{-0.2cm}
	\begin{center}
		\begingroup
		\renewcommand{\arraystretch}{1}
		\begin{tabular}{p{5.5cm}p{2.5cm}}
			\toprule
			Configuration Attribute & Attribute Value \\
			\hline
			Number of SN nodes      & 100             \\
			Computing resources     & U[50, 100]      \\
			Bandwidth resources     & U[50, 100]      \\
			ISA                     & U[1, 3]         \\
			\midrule
			VNRs                    & 2000            \\
			VNR nodes               & U[2, 10]        \\
			Computing requirements  & U[0, 50]        \\
			
			Bandwidth requirements  & U[0, 50]        \\
			ISR                     & U[1, 3]         \\
			\midrule
			VNRs for training       & 1000            \\
			VNRs for testing        & 1000            \\ 
                \bottomrule            
		\end{tabular}
		\endgroup
		\label{tab:simulation_environment_parameters}
	\end{center}
\end{table}

\begin{table}[!htb]
	\caption{parameters of the DRL-based learning agent}
	\begin{center}
		\begingroup
		\renewcommand{\arraystretch}{1}
		\begin{tabular}{p{5.5cm}p{2.5cm}}\toprule		
			Parameters                          & Value     \\ 
                \midrule
			Learning rate $(\alpha)$            & 0.001     \\
			Discount factor $(\gamma)$          & 0.998     \\
			Batch Size $(batchSize)$            & 64        \\
			Number of epoch $(epochNumber)$     & 100       \\
                \bottomrule
		\end{tabular}
		\endgroup
		\label{tab:drl_hyperparameters}
	\end{center}
  \vspace{-0.5cm}
\end{table}

\begin{figure}[!htb]
\vspace{-0.1cm}
	\centering
	\includegraphics[width=0.4\textwidth, angle=0]{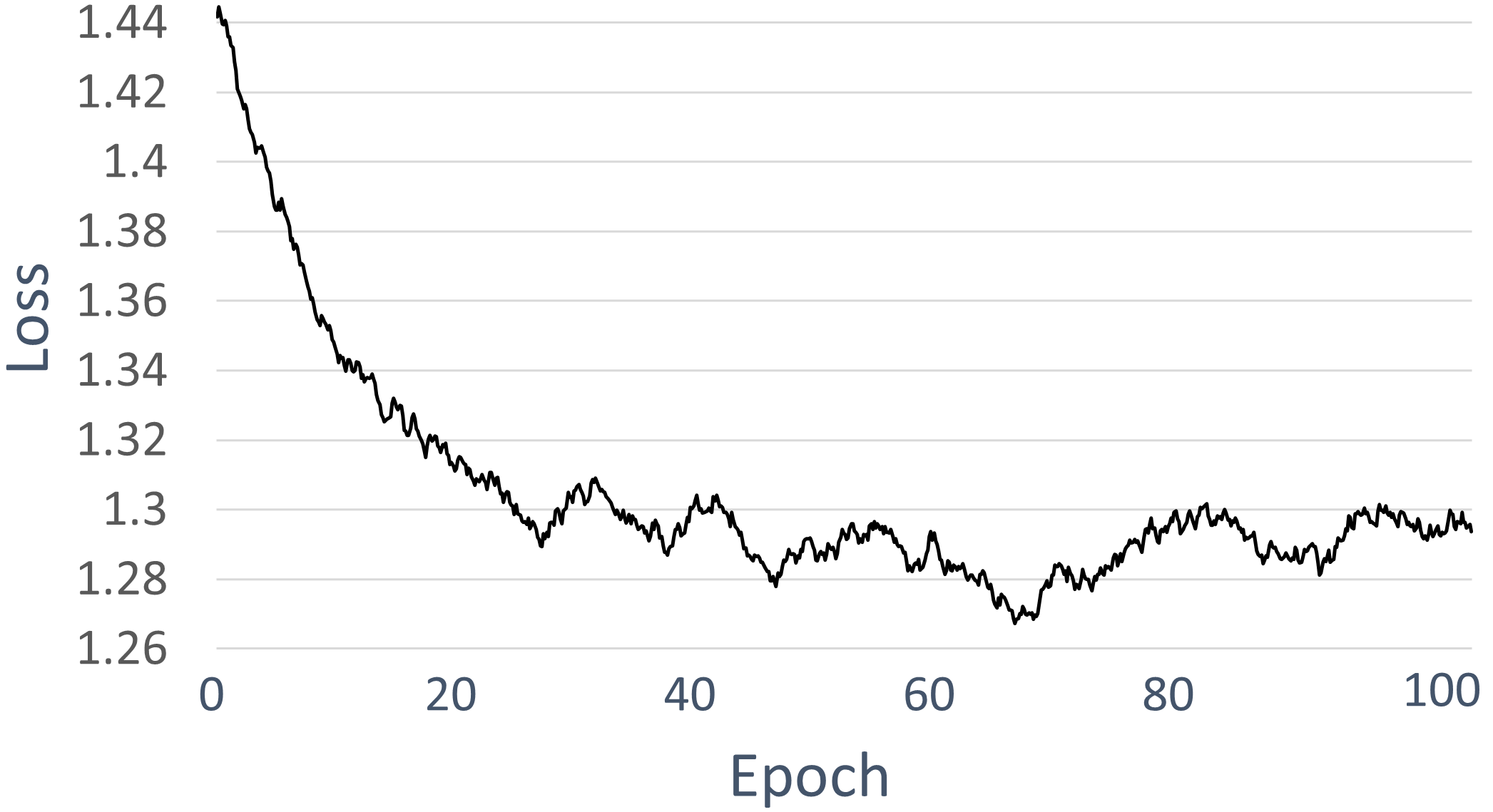}
	\caption{The loss variation during the training process.}
	\label{fig:loss}
\end{figure}

\begin{figure*}
\subfigure[\hspace*{-3.4em}]{\includegraphics[width=0.325\textwidth]{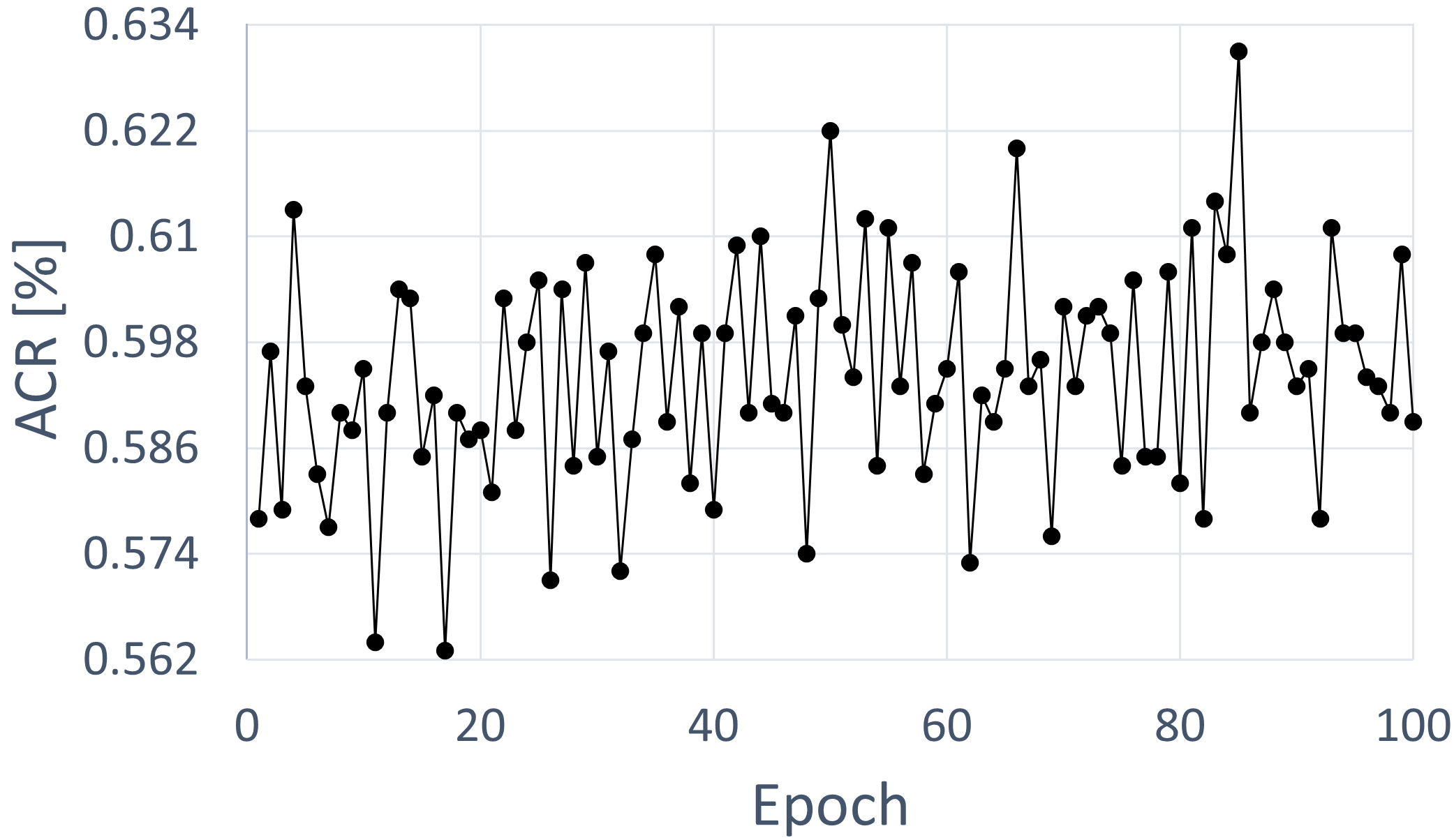}}     \subfigure[\hspace*{-3.4em}]{\includegraphics[width=0.325\textwidth]{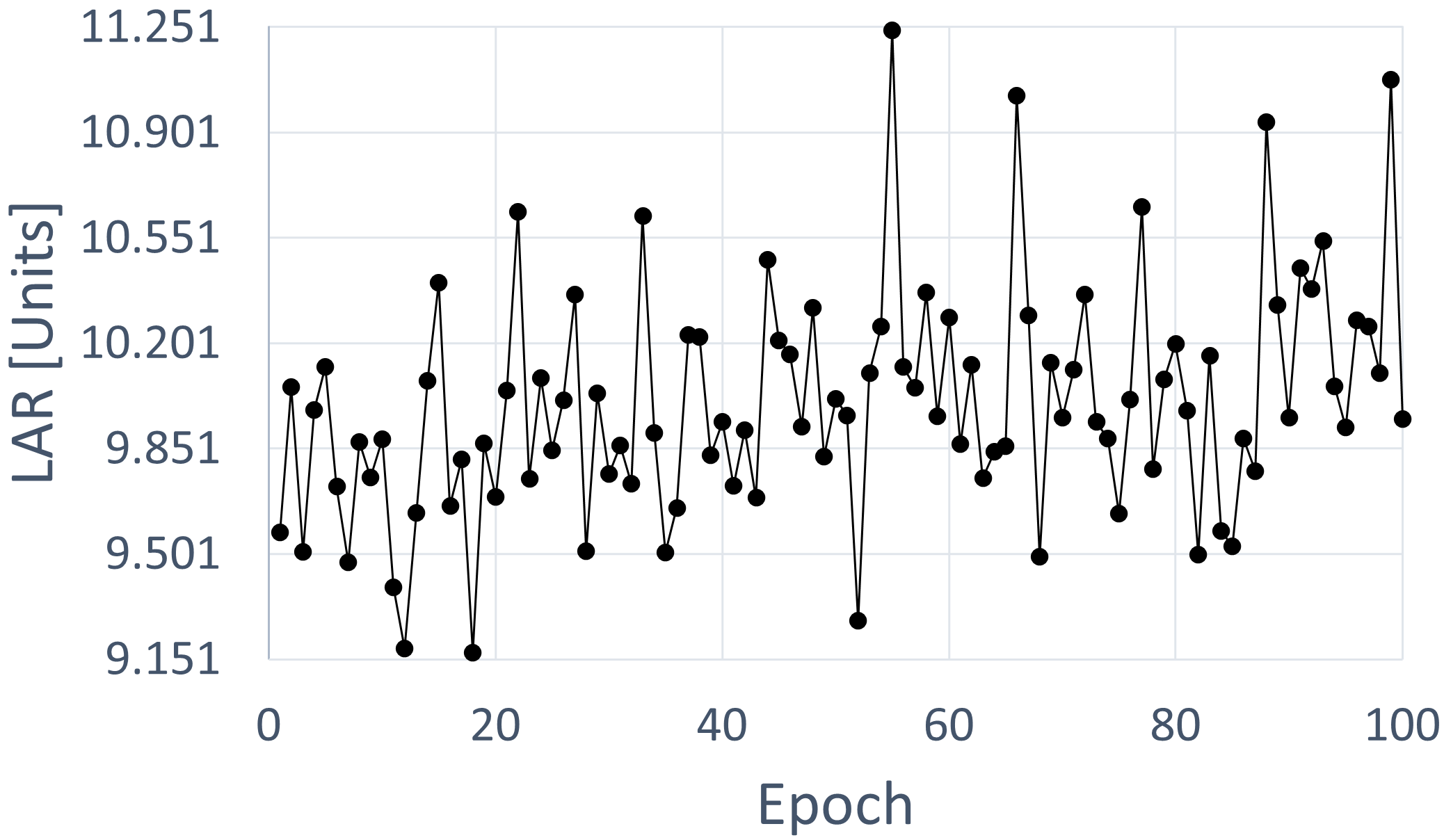}} 
\subfigure[\hspace*{-3.4em}]{\includegraphics[width=0.325\textwidth]{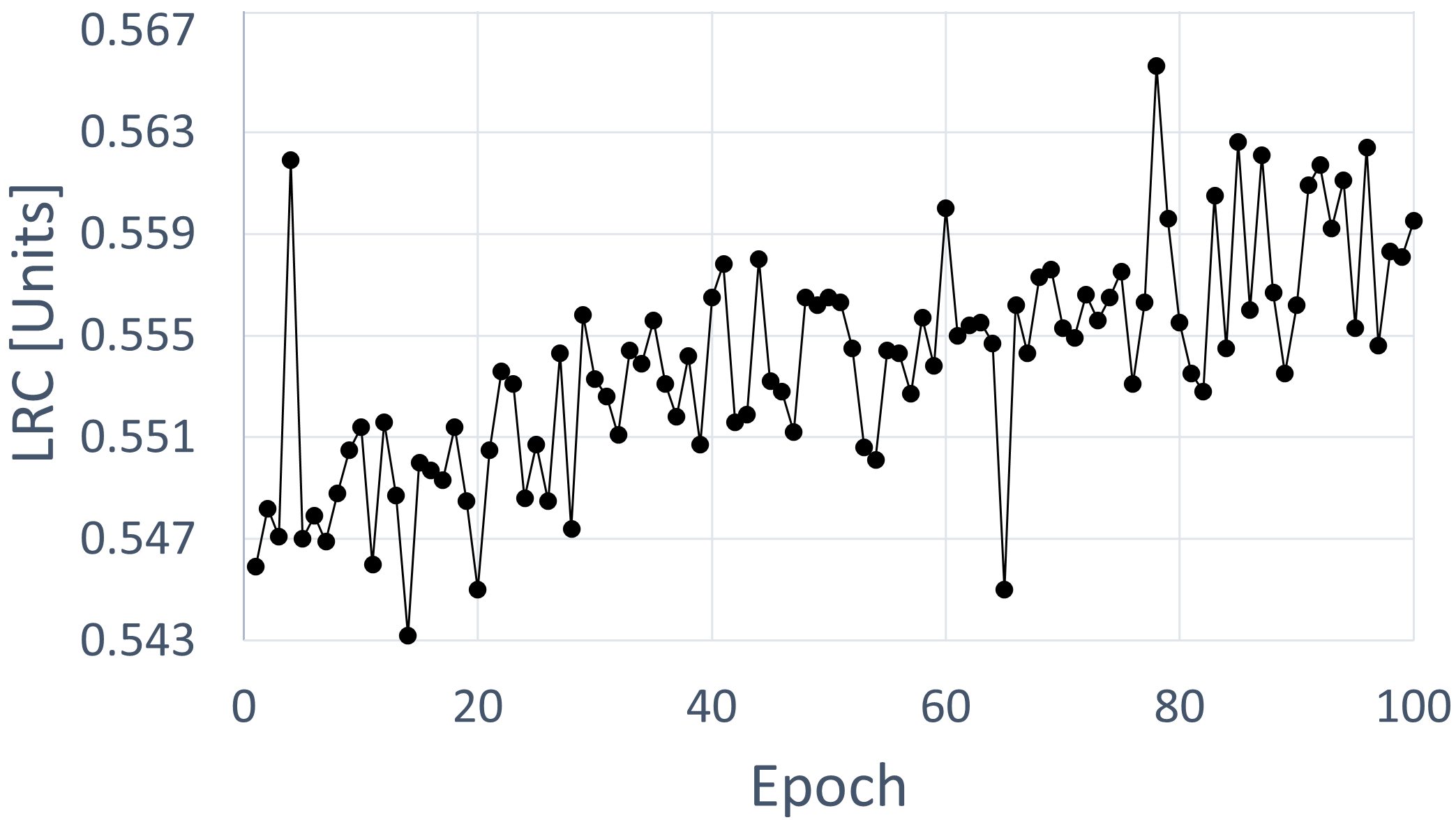}}
    \caption{Training results. (a) Performance of the VNRs acceptance ratio (ACR) on the training set. (b) Performance of long-term average revenue (LAR) on the training set. (c)   Performance of long-term revenue to cost ratio (LRC) on the training set.  }
    \label{fig:training}
    \vspace{-0.3cm}
\end{figure*}
\section{Evaluation}
\label{evaluation}
In this section, we first describe the evaluation parameters used to generate SN topology, VNRs, and comparison algorithms.
The main simulation parameters are also presented in Table.~\ref{tab:simulation_environment_parameters}.

\subsection{Evaluation Settings}
\subsubsection{Generating SN}

Using the GT-ITM tool, we create an SN that is roughly the size of a middle-sized ISP and has 100 nodes \cite{10.1145/1971162.1971168} \cite{6847918} \cite{LIU2015151} and about 500 links \cite{10.1145/1971162.1971168} \cite{LIU2015151}. The computing capacity of every SN node and the bandwidth of every SN node link is a positive real number that follows a uniform distribution and is between 50 and 100 \cite{5951812} \cite{10.1145/1971162.1971168} \cite{6847918} \cite{LIU2015151}.
\subsubsection{Generating VNRs}

The VNR arrive with an average of 5 VNRs per 100-time units according to a Poisson distribution, and they stay for the duration with an average of 500-time units according to an exponential distribution \cite{10.1145/1971162.1971168} \cite{6847918} \cite{LIU2015151}. There are 2–10 virtual nodes in each VNR \cite{5951812}. Every virtual node has a uniformly distributed computing, and bandwidth requirement between 0 and 50 units \cite{10.1145/1971162.1971168} \cite{6847918} \cite{LIU2015151}. 2000 VNRs are generated and are equally divided into two sets: a training set and a evaluation set.

\subsubsection{Comparison Algorithms}
Our algorithm is written in Python, and we use Pycharm as an IDE. For comparison, we chose the following algorithms:
NodeRank \cite{10.1145/1971162.1971168}: Computes ranking values based on RW and sorts SN nodes based on their importance. GRC \cite{6847918}: Considers global resource capacity to map VNFs onto SN nodes. MCTS \cite{7859375}: RL-based algorithm that uses Monte Carlo Tree Search (MCTS) to make the VNF placement decision. For link mapping, the shortest path algorithm is used by all evaluation algorithms.

\subsubsection{ISO-DRL\_VNE algorithm}

TensorFlow is used to build the policy network, and Xavier \cite{guo_kunin_2019} is used to initialize the policy network parameters using the normal distribution. The gradient of each training iteration is calculated using a stochastic gradient descent optimizer with a learning rate of 0.001 over 100 epochs. The key parameters of the DRL-based learning agent to reproduce our results are listed in Table.~\ref{tab:drl_hyperparameters}.

\subsubsection{ISO-DRL\_VNE Complexity}
The algorithm complexity of the ISO-DRL\_VNE algorithm is expressed as a function of the number of incoming VNRs, $C_{VNR}$, the number of substrate nodes, $C_{n^{s}}$, the dimension of the feature matrix, $d$, and the number of successfully embedded virtual nodes, $C_{n^{v}}$, and virtual links, $C_{l^{v}}$. The general form of the algorithm complexity is $O\left( C_{VNR}\left( C_{n^{s}} \cdot d + C_{n^{v}} + C_{l^{v}}\right) \right)$. This means that the algorithm's running time will increase linearly with the number of incoming VNRs, as well as with the size of the substrate network, the dimension of the feature matrix, and the number of successfully embedded virtual nodes and virtual links.

\subsection{Training Results}

Fig.\ref{fig:loss} depicts the performance of the DRL based VNE algorithm (DRL-ISO\_VNE), which shows the variation of the loss of the policy network with increasing number of training requests. The loss reaches a minimum value, a local optimum, after about 50 epochs (400 VNRs). This implies that the approximation utilized in the algorithm is effective. At the beginning of the training process, the loss value is relatively high due to the random initial parameter values, but as training continues, the learning algorithm updates the Artificial Neural Network (ANN) based on the reward signal. The decreasing trend in the loss suggests that the loss approximation performance of the DRL-based VNE algorithm is satisfactory.

Fig.\ref{fig:training} shows the performance of ISO-DRL\_VNE during the training process. Initially, the parameters of the policy network are randomly initialized and it takes time for the algorithm to adapt to the environment, causing the performance to be unstable. However, as training continues, the agent becomes more familiar with the environment and starts to take actions that lead to a larger reward, causing the performance to become more stable. By the end of the training period, the performance has reached its limit and the policy network has fully adapted to the learning environment. 

\subsection{Evaluation Results}

The comparison algorithms are heuristic, and they do not consider the long-term impact of resources and isolation constraints when allocating VNRs.
 
\begin{figure}[!htb]
	\centering
	\includegraphics[width=0.49\textwidth, angle=0]{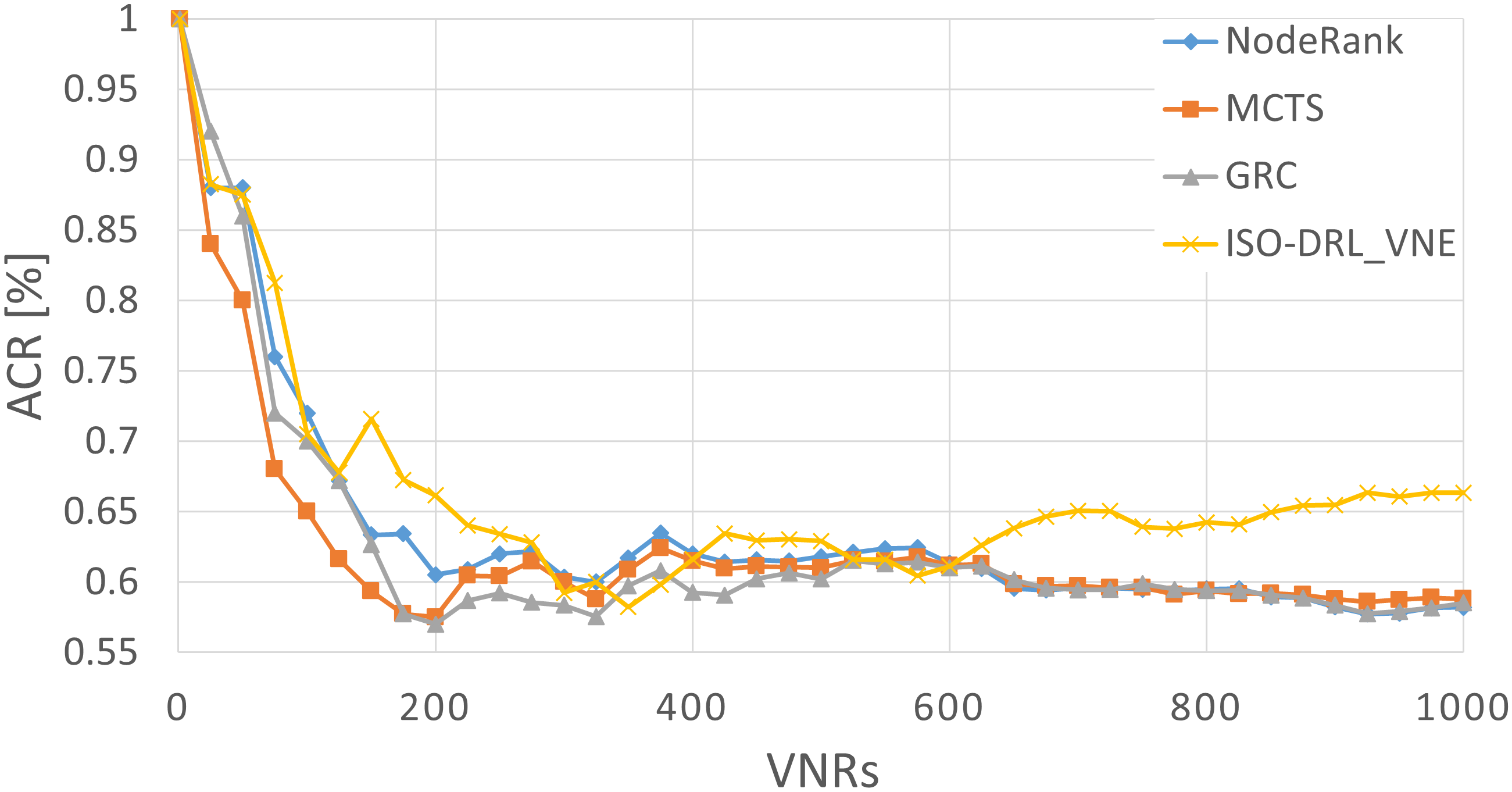}
	\caption{Acceptance ratio (ACR)}
	\label{fig:acr}
 \vspace{4.5cm}
\end{figure}

Figure \ref{fig:acr} shows the acceptance ratio (ACR) of all algorithms. The ACR is high initially because, as time passes, the SN nodes become increasingly congested, making it more difficult to allocate additional VNRs. ISO-DRL\_VNE achieves the highest acceptance rate of 0.66, up to 4.3\%,	6.4\%, and 6.0 better
than NodeRank, MCTS and GRC. 

\begin{figure}[!htb]
	\centering
	\includegraphics[width=0.49\textwidth, angle=0]{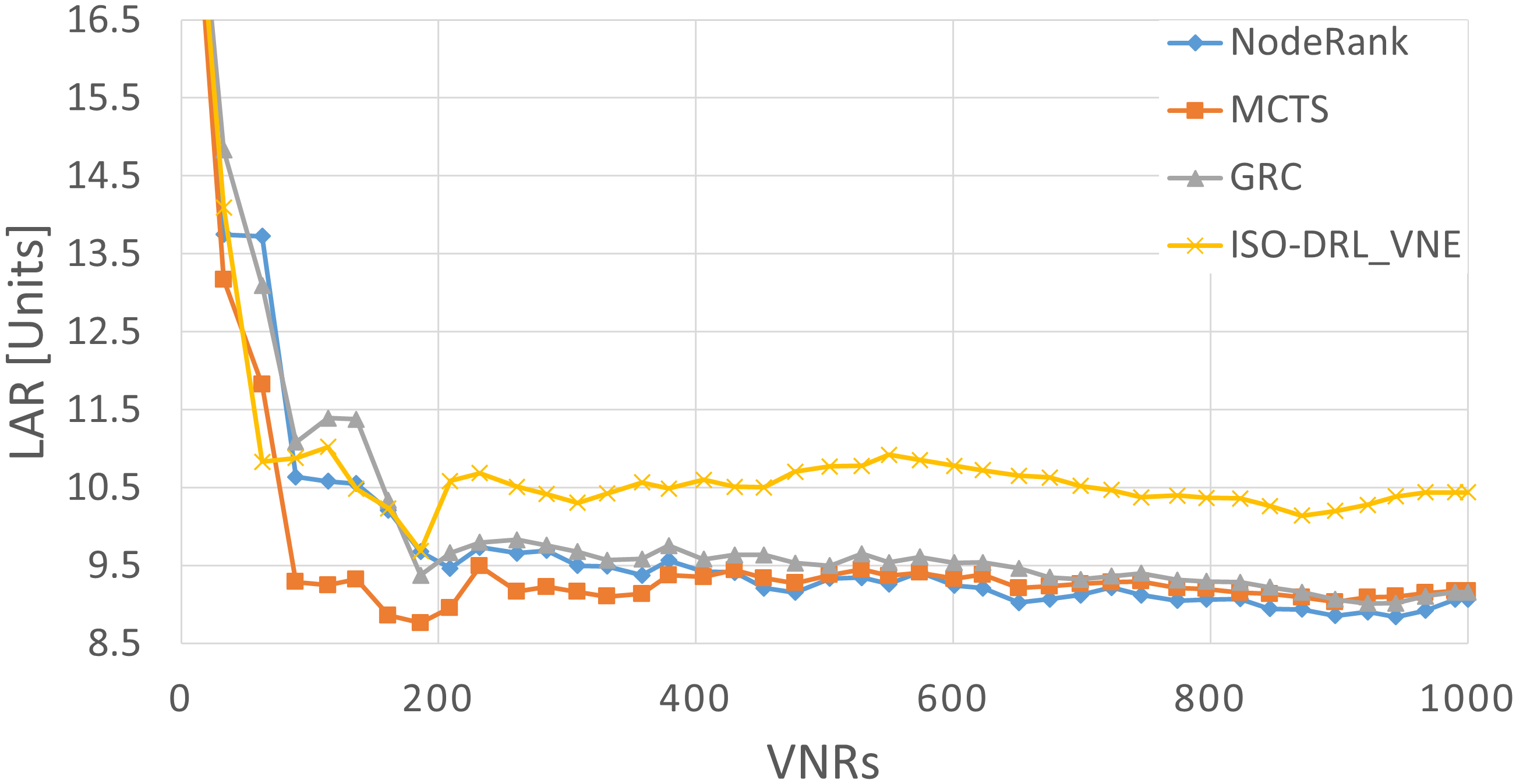}
	\caption{Long-term average revenue (LAR)}
	\label{fig:lar}
 \vspace{-0.3cm}
\end{figure}

Figure\ref{fig:lar} shows the long-term average revenue (LAR), which declines over time and has a similar trend as in Figure \ref{fig:acr}. This is because both metrics are affected by the number of resources available in SN. The ACR is high before 200 VNRs due to the abundance of SN resources that can satisfy more VNR requirements. However, due to the continued use of SN resources in the later period, the trend began to decline gradually. Overall, ISO-DRL\_VNE has a LAR of 10.6, up to 10\%, 13\%, and 7\% better than NodeRank, MCTS, and GRC. 
 
\begin{figure}[!htb]
	\centering
	\includegraphics[width=0.49\textwidth, angle=0]{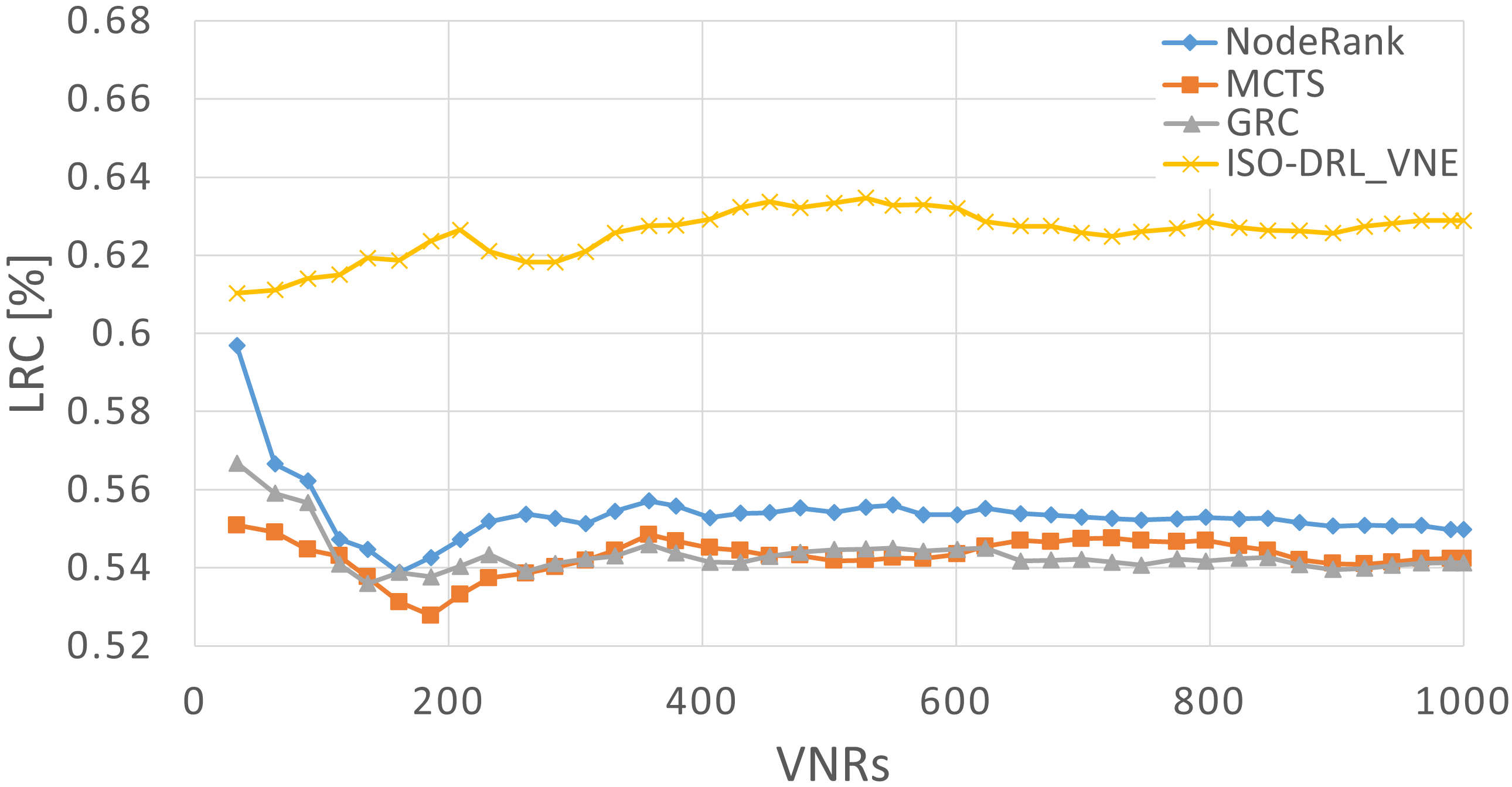}
	\caption{Long-term average revenue to cost ratio (LRC)}
	\label{fig:lrc}
  \vspace{-0.5cm}
\end{figure}

Figure \ref{fig:lrc} shows the long-term revenue-to-cost ratio (LRC). An upward trend is observed because the LRC is independent of available SN resources. This evaluation metric depends on the efficiency of the algorithm. The LRC of the ISO-DRL\_VNE algorithm is 13\%, 	15\%, and 15.3\% higher than NodeRank, MCTS, and GRC. 

The run time of ISO-DRL\_VNE algorithm is 29\%, 9\%, and 52\% lower than NodeRank, MCTS, and GRC. 

\vspace{-4.5cm}

\section{Conclusion}
\label{conclusion}

The VNE problem with an isolation levels is introduced and investigated in this study. First, this paper explains why VN VNFs need different levels of isolation requirements and proposes the VNE problem with isolation constraints. Second, to maximize the InP long-term average revenue (LAR) and long-term average revenue-to-cost ratio (LRC), an isolation-aware VNE problem is presented.  
ISO-DRL\_VNE, a DRL-based algorithm, is proposed and evaluated in comparison to three state-of-the-art algorithms. Based on the outcomes of the experiments, we found that the ISO-DRL\_VNE could enhance VN isolation while offering a good VNE strategy with high performance and practical applicability in online scenarios.

\bibliographystyle{IEEEtran}
\bibliography{bib/bibliography}

\end{document}